\documentclass[letterpaper]{elsart5p}
\journal{Physics Letters A}
\pdfoutput=1
\usepackage{amsmath}
\usepackage{amsfonts}
\usepackage{amssymb}
\usepackage{graphicx}
\renewcommand{\Re}{\mathrm{Re}}
\renewcommand{\Im}{\mathrm{Im}}

\begin{document}

\begin{frontmatter}

\title{Dynamics of a network of quadratic integrate-and-fire neurons with  bimodal heterogeneity}

\author{Viktoras Pyragas and Kestutis Pyragas}
\address{Center for Physical Sciences and Technology, Sauletekio al. 3, LT-10257 Vilnius, Lithuania}

\begin{abstract}
An exact low-dimensional system of mean-field equations for an infinite-size network of pulse coupled integrate-and-fire neurons with a bimodal distribution of an excitability parameter is derived. Bifurcation analysis of these equations shows a rich variety of dynamic modes that do not exist with a unimodal distribution of the excitability parameter. New modes include multistable equilibrium states with different levels of the spiking rate, collective oscillations and chaos. All oscillatory modes coexist with stable equilibrium states. The mean field equations are a good approximation to the solutions of a microscopic model consisting of several thousand neurons.
\end{abstract}


\end{frontmatter}

\section{\label{sec:Introduction} Introduction}
Dynamic processes in large networks of interacting neurons are the focus of intense research. Examples of collective behavior found in such systems are synchronization and collective oscillations \cite{Mirollo1990,Crook1998,Ratas2016,Ratas2019}, multistability  \cite{Ratas2016,Ratas2019,Renart2003,Montbrio2015,So2014}, collective irregular dynamics and chaos~\cite{So2014,Olmi2010,Pazo2016,Ratas2017,Ratas2018,Politi2018,Li2019}, spatiotemporal patterns~\cite{Ratas2012,Ma2013,Scholl2016,Laing2020} and others. At the microscopic level, neural networks are described by a huge number of differential equations, and their solution requires significant computer resources.  Obtaining reduced models to describe the collective dynamics of large neural populations in terms of low-dimensional dynamical systems for averaged variables is an actual problem in theoretical neuroscience. Such models are not only useful for reducing the complexity of computations, but also provide a better understanding of the mechanisms of the emergence of various dynamic modes in neural networks. Although phenomenological low-dimensional models (the so-called neural mass models) have been developed for a long time~\cite{Wilson1973,Destexhe2009}, significant advances in this direction have been achieved only recently~\cite{Schwalger2019,Coombers2019,Bick2020}. The new approach is based on the consideration of synchronizing systems by methods of statistical physics~
\cite{Gupta2018}. This approach makes it possible to derive an exact low-dimensional system of mean-field equations from the microscopic dynamics of individual neurons. Unlike phenomenological neural mass models~\cite{Wilson1973,Destexhe2009}, such equations were called the next-generation models~\cite{Coombers2019}. 

The next-generation models are based on the general mathematical method originally developed by Ott and Antonsen~\cite{Ott2008}. Analyzing the microscopic model equations of globally coupled heterogeneous phase oscillators (Kuramoto model), they discovered an ansatz that allowed them to reduce these equations to a low-dimensional system that accurately describe the macroscopic dynamics in the infinite-size (thermodynamic) limit. Later, this approach was extended to  heterogeneous neural networks composed of all-to-all pulse-coupled quadratic integrate-and-fire (QIF) neurons~\cite{Montbrio2015}, which are the normal form of class I neurons~\cite{izhi07}. The reduction procedure in Ref~\cite{Montbrio2015} was based on the Lorentzian ansatz (LA), which is different from but closely related to the Ott-Antonsen ansatz~\cite{Ott2008}. The advantage of the LA  is that it directly leads to mean-field equations for biophysically relevant macroscopic quantities: the firing rate and the mean membrane potential. This approach has been further developed in recent publications to analyze the occurrence of synchronized macroscopic oscillations in networks of QIF neurons  with a realistic synaptic coupling~\cite{Ratas2016}, in the presence of a delay in couplings~\cite{Pazo2016,Ratas2018,Devalle2017}, in the presence of noise~\cite{Ratas2019}, in the case of additional electrical coupling~\cite{Montbrio2020} and in the case of two interacting populations~\cite{Ratas2017,Segneri2020,Pyragas2021}. 

The heterogeneity in the QIF neuron networks is determined  by a distribution  function $g(\eta)$ of an internal excitability parameter $\eta$. So far, only  unimodal bell-shaped distributions, symmetric about a maximum have been considered. The reduction turns out to be especially effective when $g(\eta)$ is the Lorentzian (Cauchy) distribution. For such a distribution, the residue method allows one to reduce the network dynamics to just one equation for a complex order parameter. When QIF neurons interact via instantaneous Dirac delta pulses, such heterogeneity cannot provide macroscopic oscillations~\cite{Montbrio2015}. Oscillations were also not detected when replacing the Lorentzian distribution with a unimodal Gaussian distribution~\cite{Montbrio2015,Klinshov2021}. It is natural to ask how these results change if other forms of heterogeneity are considered. In this paper, we will address this question for the simplest choice of a nonunimodal distribution: we consider a distribution $g(\eta)$ with two peaks constructed from a linear combination of two Lorentz functions.  We found that this modification to the original problem introduces qualitatively new behaviors, including synchronized limit cycle oscillations and chaos. Note that a similar problem for a network of Kuramoto oscillators with a bimodal distribution of natural frequencies was considered in Ref.~\cite{Martens2009}.

The paper is organized as follows. Section~\ref{sec:model} describes a microscopic model of a population of pulse-coupled QIF neurons with bimodal heterogeneity. In Sec.~\ref{sec:Mean-field}, we derive the reduced mean-field equations for this model. Section~\ref{sec:Bifurcation} is devoted to the bifurcation analysis of the mean-field equations. In Sec.~\ref{sec:Microscopic}, we compare the solutions of the mean-field equations and the microscopic model. The conclusions are presented in Sec.~\ref{sec:Conclusions}.

\section{Model}
\label{sec:model}

\begin{figure}
\centering\includegraphics{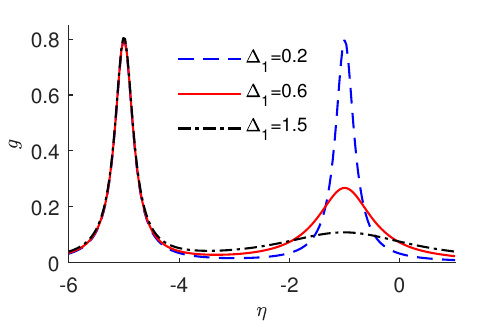}
\caption{\label{fig:g_eta} Bimodal distributions of the excitability parameter. The function $g(\eta)$ is plotted by Eqs.~\eqref{eq:g_eta} and \eqref{eq:two_lor} for fixed parameters  $\bar{\eta}_1=-1$, $\bar{\eta}_2=-5$, $\Delta_2=0.2$, $\alpha=0.5$, and different values of the parameter $\Delta_1$, which determines the half-width of the first Lorentz function in the combined distribution.}
\end{figure}

We consider a heterogeneous network of quadratic integrate-and-fire neurons, which are the canonical representatives for  class I neurons near the spiking threshold~\cite{izhi07}.  The microscopic state of the network is determined by the set of $N$ neurons' membrane potentials  $\{V_j \}_{j=1,\ldots,N}$, which satisfy the following system of $N$ ordinary differential equations (ODEs) \cite{ermentrout10}: 
\begin{equation}
\tau_m\dot{V}_{j}=  V_{j}^{2}+\eta_j+Js(t),\;\;
 \text{if} \; V_j \ge V_p \; \text{then} \; V_j \leftarrow V_{r}. \label{model}
\end{equation}
Here, $\tau_m$ is the membrane time constant,  $\eta_j$ is
a heterogeneous excitability parameter that specifies the behavior of uncoupled neurons and the term $Js(t)$ stands for the synaptic coupling, where $J$ is the synaptic weight and $s(t)$ is the normalized output signal of the network. For $J=0$, the  neurons with the negative value of the parameter $\eta_j<0$ are at rest, while the neurons with the positive value of the parameter $\eta_j>0$ generate spikes. Each time a potential $V_j$ reaches the threshold value $V_p$, it is reset to the value $V_r$, and the neuron emits an instantaneous spike which contributes to the network output:
\begin{equation}\label{I_syn}
    s(t)=\frac{\tau_m}{N}\sum_{j=1}^{N}\sum_{k \setminus t^{k}_{j}<t}\delta(t-t^{k}_{j}),
\end{equation}
where $t^{k}_{j}$ is the moment of the $k$-th spike of the $j$-th
neuron and $\delta(t)$ is the Dirac delta function. We use the multiplier $\tau_m$ in the definition of $s(t)$ in order to get the dimensionless output. 
Because of the quadratic nonlinearity, $V_j$ reaches infinity in a finite time, and this allows us to choose the threshold parameters as $V_{p}= -V_{r} = \infty$. With this choice, the QIF neuron can be transformed into a theta neuron. This choice is also crucial for the derivation of the reduced mean-field equations in an infinite  size limit $N \to \infty$ \cite{Montbrio2015}. The  mean-field equations are usually derived under the assumption that the values of the parameter $\eta$ are distributed according to the Lorentzian density function. For such a unimodal distribution, the reduction method is especially effective, since it leads to only two equations for two order parameters. Here we consider the case of bimodal distributions defined by the linear combination
\begin{equation}\label{eq:g_eta}
    g(\eta)=\alpha_{1}g_{1}(\eta)+\alpha_{2}g_{2}(\eta)\equiv\alpha g_{1}(\eta)+(1-\alpha)g_{2}(\eta)
\end{equation}
of two Lorentzian distributions  
\begin{equation}\label{eq:two_lor}
    g_{1,2}(\eta)=\frac{1}{\pi}\frac{\Delta_{1,2}}{(\eta-\bar{\eta}_{1,2})^{2}+\Delta^{2}_{1,2}},
\end{equation}
generally characterized by different values of the half-widths $\Delta_{1,2}$ and the centers $\bar{\eta}_{1,2}$. The weight parameters  $\alpha_{1}$ and $\alpha_{2}$ satisfy the normalization condition $\alpha_{1}+\alpha_{2}=1$,  which makes it possible to express them in terms of one parameter $\alpha$:   $\alpha_{1}=\alpha$ and $\alpha_{2}=1-\alpha$. Examples of distributions $g(\eta)$ obtained from Eqs.~\eqref{eq:g_eta} and \eqref{eq:two_lor} for fixed $\bar{\eta}_1=-1$, $\bar{\eta}_2=-5$, $\Delta_2=0.2$, $\alpha=0.5$, and different values of the parameter $\Delta_1$ are shown in Fig.~\ref{fig:g_eta}. Below, we show that the microscopic dynamics of an infinite size network with a two-Lorentzian distribution Eq.~\eqref{eq:g_eta} reduces to four exact ODEs for four order parameters. We will analyze solutions of these equations depending on the coupling strength $J$ and the parameter  $\Delta_1$, which determines the half-width of the first Lorentz function in the combined distribution Eq.~\eqref{eq:g_eta}.

\section{Mean-field equations}
\label{sec:Mean-field}
 
The system~\eqref{model} can be reduced using the LA method~\cite{Montbrio2015}. This method is usually applied to QIF neural networks with a unimodal distribution $g(\eta)$  described by a single Lorentz function. Here we will briefly reproduce this method for the case of a bimodal distribution defined by Eqs.~\eqref{eq:g_eta} and \eqref{eq:two_lor}.
 
In the thermodynamic limit $N \to \infty$, the  population state can be characterized by the density function $\rho(V|\eta,t)$, which evolves according to the continuity equation
\begin{equation}\label{cont_eq}
   \tau_m \partial_{t}\rho+\partial_{V}[(V^{2}+\eta+Js)\rho]=0.
\end{equation}
According to the LA theory~\cite{Montbrio2015}, solutions of Eq.~\eqref{cont_eq} generically (independently of the initial conditions) converge to a Lorentzian-shaped function
\begin{equation}\label{rho_LA}
    \rho(V|\eta,t)=\frac{1}{\pi}\frac{x(\eta,t)}{[V-y(\eta,t)]^{2}+x^{2}(\eta,t)}
\end{equation}
with two time-dependent variables, $x(\eta,t)$ and $y(\eta,t)$, which
define the half-width and the center of the voltage distribution of neurons with a given $\eta$. The LA ansatz~\eqref{rho_LA}  allows us  to reduce a partial differential Eq.~\eqref{cont_eq}  to an ODE:
\begin{equation}\label{w_eta}
    \tau_m \partial_{t}w(\eta,t) = i [ \eta+Js(t)-w^{2}(\eta,t)],
\end{equation}
where $w(\eta,t)=x(\eta,t)+i y(\eta,t)$ is a complex variable. The variables $x(\eta,t)$ and $y(\eta,t)$ have clear physical meanings. For a fixed $\eta$, the neurons firing rate $R(\eta,t)$ is related to the Lorentzian half-width by $R(\eta,t) = x(\eta,t)/(\pi\tau_m)$. This relation is obtained by estimating the probability flux $R(\eta,t) = \rho(V_p|\eta,t) \dot{V}(V_p|\eta,t)$ through the threshold $V_p=\infty$. Below we will use a dimensionless firing rate $r(\eta,t)=\tau_m R(\eta,t)=\Re [w(\eta,t)]/\pi$. Then the averaged   value of the dimensionless firing rate $r(\eta,t)$ over $\eta$,
\begin{equation}\label{r_t}
    r(t)=\frac{1}{\pi}\Re\int^{+\infty}_{-\infty} w(\eta,t)g(\eta)d\eta,
\end{equation}
is equal to the network output $s(t)$. The equality $s(t)=r(t)$ turns the Eqs.~\eqref{w_eta} and \eqref{r_t} into a closed system of integro-differential equations. In these equations, the averaged value of the variable $y(\eta, t)=\Im [w(\eta, t)]$ over $\eta$ determines the mean membrane potential
\begin{equation}\label{v_t}
    v(t)=\Im \int^{+\infty}_{-\infty}w(\eta,t)g(\eta)d\eta.
\end{equation}
For the distribution $g(\eta)$ in the form of Eqs.~\eqref{eq:g_eta} and \eqref{eq:two_lor}, the integrals in Eqs.~\eqref{r_t} and \eqref{v_t} can be
evaluated using the residue theory. Namely, function $w(\eta,t)$ is analytically continued into a complex-valued $\eta$, and the
integration contour is closed in the lower half-plane. Expanding $g(\eta)$ in partial fractions as
\begin{eqnarray}\label{g_eta_c}
    g(\eta)&=&\frac{1}{2\pi i} \Big[ \frac{\alpha_{1}}{(\eta-\bar{\eta}_{1})-i\Delta_{1}}-\frac{\alpha_{1}}{(\eta-\bar{\eta}_{1})+i\Delta_{1}}\nonumber \\
    &+& \frac{\alpha_{2}}{(\eta-\bar{\eta}_{2})-i\Delta_{2}}-\frac{\alpha_{2}}{(\eta-\bar{\eta}_{2})+i\Delta_{2}} \Big],
\end{eqnarray}
we find it has four simple poles at $\eta=\bar{\eta}_1 \pm i\Delta_1$ and $\eta=\bar{\eta}_2 \pm i\Delta_2$. Since the values of the integrals \eqref{r_t} and \eqref{v_t} are determined by the poles $\eta=\bar{\eta}_{1,2} - i\Delta_{1,2}$ of $g(\eta)$ in the lower half  plane, we obtain 
\begin{subequations}\label{eq_rv}
\begin{eqnarray}
r(t)&=& \Re \left[\alpha_1 w_1(t)+\alpha_2 w_2(t)\right]/\pi,\\
v(t)&=&\Im \left[\alpha_1 w_1(t)+\alpha_2 w_2(t)\right],
 \end{eqnarray}
\end{subequations}
where $w_{1,2}(t)\equiv w(\bar{\eta}_{1,2}-i\Delta_{1,2},t)$ are two complex order parameters. With the help of these parameters, the integro-differential system of Eqs.~\eqref{w_eta} and \eqref{r_t} turns into a closed system of two complex ODEs:
\begin{subequations}\label{eq_w12}
\begin{eqnarray}
\tau_m\dot{w}_1(t)&=& i [ \bar{\eta}_1-i\Delta_1+Jr(t)-w_1^{2}(t)],\\
\tau_m\dot{w}_2(t)&=& i [ \bar{\eta}_2-i\Delta_2+Jr(t)-w_2^{2}(t)].
 \end{eqnarray}
\end{subequations}
To rewrite this system in a real valued form, we define the real and imaginary parts of the two complex order parameters as $w_{1,2}(t)=\pi r_{1,2}(t)+iv_{1,2}(t)$. Substituting these expressions into Eqs.~\eqref{eq_w12} and separating the real and imaginary parts, we get the closed system of four ODEs for four real order parameters:
\begin{subequations}\label{eq_r12v12}
\begin{eqnarray}
\tau_m\dot{r}_1(t)&=& \Delta_1/\pi+2r_1(t)v_1(t),\\
\tau_m\dot{v}_1(t)&=& \bar{\eta}_1+Jr(t)-\pi^2 r_1^2(t)+v_1^{2}(t),\\
\tau_m\dot{r}_2(t)&=& \Delta_2/\pi+2r_2(t)v_2(t),\\
\tau_m\dot{v}_2(t)&=& \bar{\eta}_2+Jr(t)-\pi^2 r_2^2(t)+v_2^{2}(t).
 \end{eqnarray}
\end{subequations}
According to the Eqs.~\eqref{eq_rv}, the mean spiking rate $r(t)$ and the mean membrane potential $v(t)$ of the network are expressed through these parameters as
\begin{subequations}\label{eq_rv1}
\begin{eqnarray}
r(t)&=& \alpha_1 r_1(t)+\alpha_2 r_2(t),\\
v(t)&=& \alpha_1 v_1(t)+\alpha_2 v_2(t).
 \end{eqnarray}
\end{subequations}
Although the parameters $r_{1,2}$ and $v_{1,2}$ are formally introduced, we can give them a physical meaning. The Eqs.~\eqref{eq_r12v12} can be interpreted as mean-field equations describing two globally connected populations of QIF neurons with different numbers of  neurons $N_{1}$ and $N_{2}$ in each of the populations and with different distributions $g_{1}(\eta)$ and $g_{2}(\eta)$ of the heterogeneity parameter $\eta$, given by the Eq.~\eqref{eq:two_lor}. In this interpretation, $\alpha_{1,2}=N_{1,2}/(N_1+N_2)$ are the proportion of neurons in each subpopulation, $r_{1,2}(t)$ and $v_{1,2}(t)$ are the mean spiking rates and the mean membrane potentials in each subpopulation, respectively, and the Eqs.~\eqref{eq_rv1} determine their global mean values for the entire population.

\section{\label{sec:Bifurcation} Bifurcation analysis of the mean-field equations}

We start the analysis of the mean-field equation by determining the equilibrium points and their stability. The coordinates $(r_1^*, v_1^*, r_2^*, v_2^*)$ of equilibrium points in the four dimensional phase space of the system are obtained by equating the right-hand sides (RHS) of the Eqs.~\eqref{eq_r12v12} to zero. In the general case, this problem requires solving a system of polynomial equations. However, if we are interested in the dependence of the equilibrium points on the coupling strength $J$, we do not need to solve the polynomial equations.  In a parametric form, this dependence can be written as follows:
\begin{subequations}\label{eq_eqp}
\begin{eqnarray}
r_{1,2}^*(p)&=& \frac{1}{\sqrt{2}\pi} \sqrt{\bar{\eta}_{1,2}+p+\sqrt{(\bar{\eta}_{1,2}+p)^2+\Delta_{1,2}^2}}, \label{eq_eqpa}\\
r^*(p)&=& \alpha_1 r_1^*(p)+\alpha_2 r_2^*(p),\label{eq_eqpb}\\
J(p)&=& p/r^*(p).\label{eq_eqpc}
 \end{eqnarray}
\end{subequations}
Here $p=Jr$ is considered as an independent parameter. The Eq.~\eqref{eq_eqpa} is obtained by equating the RHS of the Eqs.~\eqref{eq_r12v12} to zero and solving them with respect to the variables $r_1$ and $r_2$. In addition, the dependences of the equilibrium values of the average potentials on the parameter $p$ have the form: $v_{1,2}^*(p)=-\Delta_{1,2}/2\pi r_{1,2}^*(p)$ and $v^*(p)= \alpha_1 v_1^*(p)+\alpha_2 v_2^*(p)$. 

\begin{figure}
\centering\includegraphics{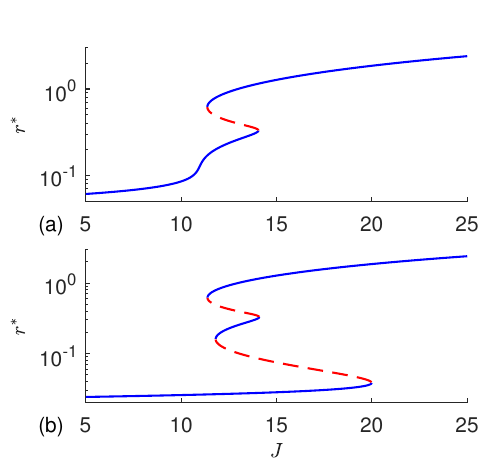}
\caption{\label{fig:Fixed_points} Multistability of equilibrium states in a network of QIF neurons with bimodal heterogeneity.  Equilibrium spiking rate $r^*$ as a function of the coupling strength $J$ is shown in semi-log coordinates for (a) $\Delta_1=0.6$ and (b) $\Delta_1=0.2$. The rest of the parameters are the same as in Fig.~\ref{fig:g_eta}. The solid blue and red dashed curves show the stable and unstable branches, respectively. Their mergers are the saddle-node bifurcation points.}
\end{figure}

The dependences of the equilibrium spiking rate $r^*$ on the coupling strength $J$, obtained from the Eqs.~\eqref{eq_eqp} for two different values of the parameter $\Delta_1$, are shown in Fig.~\ref{fig:Fixed_points}. Branches of stable equilibrium are shown by solid blue curves, and unstable ones by red dashed curves. The stability of equilibrium states was established by solving the eigenvalue problem 
\begin{equation} \label{Eigen}
\det (A-\lambda \tau_m I)=0
\end{equation} 
of the linearized system of Eqs.~\eqref{eq_r12v12}, where
\begin{equation} \label{Jacob}
A=
\begin{pmatrix}
2v_1^* & 2r_1^* & 0 & 0\\ 
J\alpha_1-2\pi^2 r_1^*& 2v_1^* & J\alpha_2 & 0\\ 
0 & 0 & 2v_2^* & 2r_2^* \\ 
J\alpha_1 & 0 & J\alpha_2-2\pi^2 r_2^* & 2v_2^*
\end{pmatrix}
\end{equation} 
is the Jacobian matrix, $I$ is the identity matrix and $\lambda$ is the eigenvalue. The equilibrium state is stable if all eigenvalues are negative, and unstable if at least one of the eigenvalues is positive. We see that at intermediate values of the coupling strength, the system is multistable. The maximum number of stable equilibrium states, characterized by different levels of the spiking rate, increases from two [Fig.~\ref{fig:Fixed_points}(a)] to three [Fig.~\ref{fig:Fixed_points}(b)] when the value of the $\Delta_1$ parameter decreases from  $0.6$  to $0.2$.

\begin{figure}
\centering\includegraphics{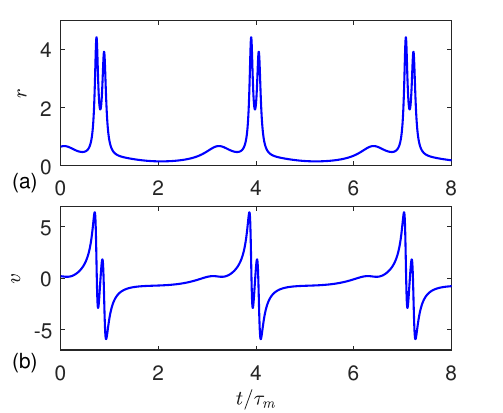}
\caption{\label{fig:LC_osc_macro}  Periodic oscillatory mode of the QIF neuron network.  The post transient dynamic of the spiking rate and the mean membrane potential shown in (a) and (b), respectively, were obtained at $\Delta_1=0.6$ and $J=16$ by solving the Eqs.~\eqref{eq_r12v12}  with zero initial conditions  $(r_1,v_1,r_2,v_2)=(0,0,0,0)$. The rest of the parameters are the same as in Fig.~\ref{fig:g_eta}.}
\end{figure}

Figure~\ref{fig:Fixed_points} shows that only saddle-node bifurcations take place at the equilibrium points. Thus the birth of a limit cycle from equilibrium is impossible here. The absence of a Hopf bifurcation reduces the likelihood of the occurrence of limit cycle oscillations in this system. Here, such oscillations can arise only through global bifurcations, and their search is nontrivial. Nevertheless, we managed to find a limit cycle coexisting with a stable equilibrium state. Figure~\ref{fig:LC_osc_macro} shows an example of limit cycle oscillations obtained for $\Delta_1=0.6$, $J=16$ and other parameters the same as in Fig.~\ref{fig:g_eta}. We found the limit cycle by solving the Eqs.~\eqref{eq_r12v12} with zero initial conditions $(r_1,v_1,r_2,v_2)=(0,0,0,0)$.

\begin{figure}
\centering\includegraphics{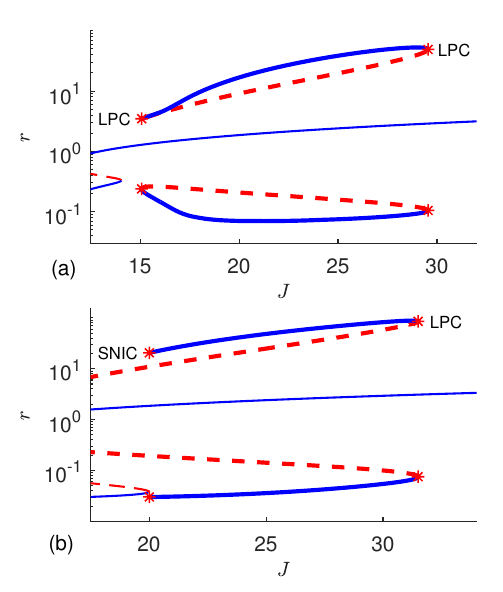}
\caption{\label{fig:bif_J_r} One-parameter bifurcation diagrams showing the evolution of limit cycles. The dependence of the firing rate $r$ on the coupling strengths $J$ for (a) $\Delta_1=0.6$ and (b) $\Delta_1=0.2$, and other parameters the same as in Fig.~\ref{fig:g_eta}. The bold blue and red dashed curves show the maximum and minimum of the stable and unstable limit cycle, respectively. The thin blue and red dashed curves show respectively the evolution of the stable and unstable equilibrium states, which are redrawn from Fig.~\ref{fig:Fixed_points}. The red asterisks marked with the letters LPC and SNIC denote the limit point of cycles bifurcation and  the saddle-node on an invariant circle bifurcation, respectively. These and other bifurcation diagrams in this paper were built using the MatCont package~\cite{matcont}.}
\end{figure}

Having a limit cycle for fixed values of parameters, we can continue it into a larger domain of parameters. Figure~\ref{fig:bif_J_r}(a) shows the continuation of the given limit cycle by changing the $J$ parameter. The bold blue and red dashed curves show the maximum and minimum of the stable and unstable limit cycle, respectively. The thin blue and red dashed curves show respectively the evolution of the stable and unstable equilibrium states. The latter are actually redrawn from Fig.~\ref{fig:Fixed_points}. We see that as $J$ increases, oscillations arise and disappear through a limit point of cycles (LPC) bifurcation. This is a global bifurcation in which two limit cycles, stable and unstable, collide and annihilate each other. The system is bistable in the entire range of oscillations: a stable limit cycle coexists with a stable state of equilibrium.

Depending on the parameter $\Delta_1$, the limit cycle can arise through another global bifurcation. Figure~\ref{fig:bif_J_r}(b) shows a one-parameter bifurcation diagram $r$ vs $J$ for $\Delta_1=0.2$ and other parameters the same as in Fig.~\ref{fig:g_eta}. Now, as $J$ increases, the limit cycle arises through the saddle-node on an invariant circle (SNIC) bifurcation. In SNIC bifurcation, a pair of fixed points on a closed curve coalesce to disappear, converting the curve to a periodic orbit with an infinite period. Here, as in Fig.~\ref{fig:bif_J_r}(a), the system is bistable in the entire oscillation range, and at large $J$ oscillations also disappear due to the LPC bifurcation.

\begin{figure}
\centering\includegraphics{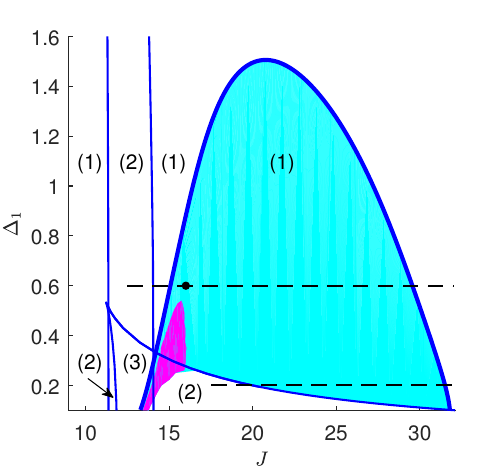}
\caption{\label{fig:bif_J_del1} Two-parameter bifurcation diagrams in the plane of parameters $(J,\Delta_1)$. Other parameters are the same as in Fig.~\ref{fig:g_eta}. Thin solid curves represent saddle-node bifurcations of equilibrium points separating regions with different numbers of stable equilibrium states. In different regions, these numbers are shown in brackets. The bold curve represents the LPC bifurcation. Oscillations occur in the colored area. Cyan indicates simple limit cycle osculations, and  magenta represents complex oscillations, including chaos. The horizontal dashed lines $\Delta_1=0.6$ and $\Delta_1=0.2$ correspond to the one-parameter bifurcation diagrams shown in Figs.~\ref{fig:bif_J_r}(a) and \ref{fig:bif_J_r}(b), respectively. The black dot denotes the values
of the $(J,\Delta_1)$ parameters used in Fig.~\ref{fig:LC_osc_macro}.}
\end{figure}

A richer scenario of dynamic modes can be seen in the two-parameter ($J,\Delta_1 $) bifurcation diagram shown in Fig.~\ref{fig:bif_J_del1}. Here, thin solid curves represent saddle-node bifurcations of equilibrium points. They separate regions with different number of stable equilibrium states. The numbers of stable equilibrium states in different regions are indicated in brackets. The colored area indicates the presence of oscillations.  At the top, the oscillations are limited by the LPC bifurcation, represented by the bold curve. At the bottom, at $J>16.7$, the oscillations are limited by the SNIC bifurcation. The cyan area denotes simple limit cycle oscillations, and the magenta area represents complex oscillations, including chaos. The horizontal dash-dotted lines show cross-sections of two-parameter bifurcation
diagram, which correspond to the one-parameter diagrams presented in Fig.~\ref{fig:bif_J_r} (see figure caption for details).

\begin{figure}
\centering\includegraphics{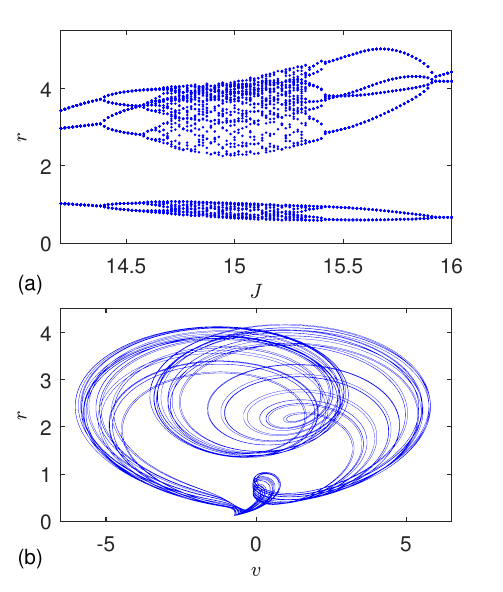}
\caption{\label{fig:chaos} An illustration of the complex network dynamics that occurs in the magenta area in Fig.~\ref{fig:bif_J_del1}. (a) One-parameter bifurcation diagram showing the dependence of the oscillation peaks of the spiking rate $r$ on the coupling strength $J$ at a fixed $\Delta_1=0.3$. (b) Chaotic oscillations in ($v,r$) coordinates at $\Delta_1=0.3$ and $J=15$. The rest of the parameters are as in Fig.~\ref{fig:bif_J_del1}.}
\end{figure}

We will now discuss the complex oscillation mode that occurs in the magenta region in Fig.~\ref{fig:bif_J_del1}. This region was identified by analyzing one-parameter bifurcation diagrams, which show the dependence of the oscillation peaks of the spiking rate $r$ on the coupling strength $J$ for different fixed values of $\Delta_1$. An example of such a diagram for a fixed $\Delta_1=0.3$ and a coupling strength varying in the interval $J \in [14.2, 16]$ is shown in Fig.~\ref{fig:chaos}(a). In the left and right parts of the interval, the system shows period doubling bifurcations, and chaotic oscillations are observed in the middle of the interval. An example of chaotic oscillations in ($v,r$) coordinates at $J=15$ is shown in Fig.~\ref{fig:chaos}(b). In this chaotic regime, the spectrum of the Lyapunov exponents of the system~\eqref{eq_r12v12} is $\lambda = \{0.13,  0,  -0.78,  -1.29\}/\tau_m$.

\section{\label{sec:Microscopic} Modeling microscopic dynamics}

The reduced mean-field Eqs.~\eqref{eq_r12v12} are obtained in the limit of a network of infinite size, while real networks consist of a finite number of neurons. A natural question arises as to how well the mean-field equations predict the behavior of finite networks. To answer this question, we simulated microscopic Eqs.~\eqref{model} for  a large number of neurons and compared  the results with solutions of the mean-field Eqs.~\eqref{eq_r12v12}.

Numerical simulation of the Eqs.~\eqref{model} is more convenient after changing the variables
\begin{equation}
V_j = \tan(\theta_j/2)\label{eq_transf_tet}
\end{equation}
that turn QIF neurons into theta neurons. Theta neurons avoid the problem associated with jumps of infinite size (from $+ \infty $ to $-\infty $) of the membrane potential $V_j$ of the QIF neuron at the moments of firing. At these moments, the phase of the theta neuron simply crosses the value  $\theta_j=\pi$. For theta neurons, the Eqs.~\eqref{model} are transformed into
\begin{equation}
\tau_m \dot{\theta}_{j} = 1-\cos \left(\theta_{j}\right)
+\left[1+\cos \left(\theta_{j}\right)\right]\left[\eta_{j}+J s(t) \right]. \label{theta_j}
\end{equation}
These equations were integrated by the Euler method with a time step of $d t = 10^{-4}\tau_m$. The values of the heterogenous parameter defined by two  Lorentzian distributions \eqref{eq:two_lor} were deterministically generated using $\eta_j=\bar{\eta}_1+\Delta_1 \tan(\pi/2)(2j-N_1-1)/(N_1+1)]$ for   $j=1,\ldots, N_1$ and $\eta_j=\bar{\eta}_2 + \Delta_2 \tan(\pi/2) (2j-N_1-N-1) / (N_2+1)]$ for   $j=N_1+1,\ldots, N$, where $N_{1,2}/N=\alpha_{1,2}$ and $N_1+N_2=N$. More information on numerical modeling of Eqs.~\eqref{theta_j} can be found in Ref.~\cite{Ratas2016}. To compare the results obtained from the microscopic model Eqs.~\eqref{theta_j} with the solutions of the reduced mean-field Eqs.~\eqref{eq_r12v12}, we calculate the Kuramoto order parameters~\cite{Kuramoto2003}
\begin{equation}
\label{eq_Z}
Z_{1}=\frac{1}{N_1}\sum\limits_{j=1}^{N_1}\exp(i \theta_j), \quad Z_{2}=\frac{1}{N_2}\sum\limits_{j=N_1+1}^{N}\exp(i \theta_j)
\end{equation} 
for each subpopulation and use their relation with the spiking rates $r_{1,2}$ and the mean membrane potentials $v_{1,2}$~\cite{Montbrio2015}: 
\begin{equation}
\label{eq_W}
r_{1,2}=\frac{1}{\pi}\operatorname{Re}\left(\frac{1-Z_{1,2}^*}{1+Z_{1,2}^*}\right), \quad v_{1,2}=\operatorname{Im}\left(\frac{1-Z_{1,2}^*}{1+Z_{1,2}^*}\right),
\end{equation} 
where $Z^*_{1,2}$ means complex conjugate of $Z_{1,2}$. Then the global means $r$ and $v$ are determined from the Eqs.~\eqref{eq_rv1}.

\begin{figure}
\centering
	\includegraphics{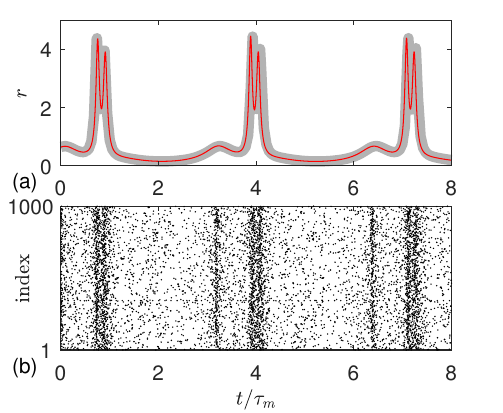}
\caption{\label{netw_dyn_micro} Dynamics of a population of $5000$ neurons in the limit cycle mode and its approximation by mean-field equations. The values of the parameters are the same as in Fig.~\ref{fig:LC_osc_macro}. (a) The
firing rate obtained from the microscopic model Eqs.~\eqref{theta_j} (thick gray curve) and from the mean-field Eqs.~\eqref{eq_r12v12} (thin red curve). (b) Raster plots of $1000$ randomly selected neurons. Here, the dots show the spike moments for each neuron, where the vertical axis indicates neuron numbers. 
}
\end{figure}

Figure ~\ref{netw_dyn_micro}(a) compares the dynamics of the microscopic model for $N=5000$ neurons and the mean-field equations. The values of the parameters correspond to the limit cycle oscillations shown in Fig.~\ref{fig:LC_osc_macro}. The time traces of the firing rate obtained from the microscopic model Eqs.~\eqref{theta_j} and the mean-field Eqs.~\eqref{eq_r12v12} are in complete agreement with each other. The network behavior on the microscopic level can be seen in raster plots shown in Fig.~\ref{netw_dyn_micro}(b).

\begin{figure}
\centering\includegraphics{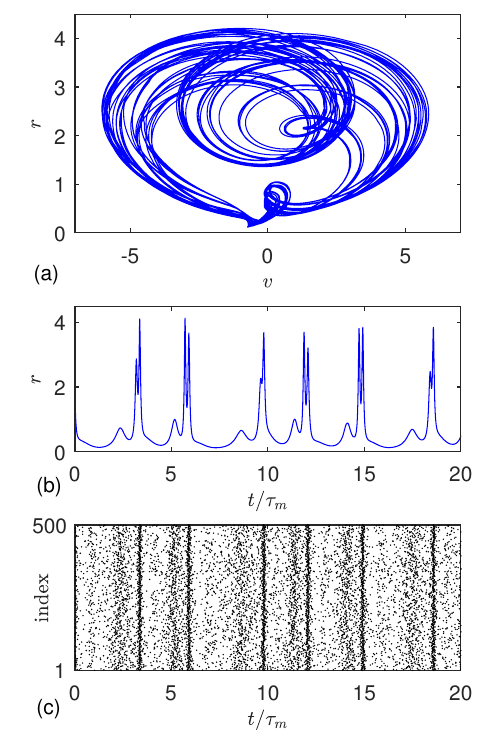}
\caption{\label{fig:chaos_micro} Chaotic network dynamics obtained from the microscopic model Eqs.~\eqref{theta_j} with $N=10^4$ neurons. (a) Phase portrait of the system in $(v,r)$ coordinates (compare with Fig.~\ref{fig:chaos}(b)). (b) Dynamics of the firing rate and (c) raster plots of $500$ randomly selected neurons. The parameter values are the same as in Fig.~\ref{fig:chaos}(b).}
\end{figure}

In Fig.~\ref{fig:chaos_micro}(a), we use the microscopic model Eqs.~\eqref{theta_j} with $N=10^4$ neurons to reproduce the chaotic mean-field dynamics shown in Fig.~\ref{fig:chaos}(b). We see that the phase portraits in Figs.~\ref{fig:chaos_micro}(a) and \ref{fig:chaos}(b) are very similar. For completeness, the dynamics of the firing rate and raster plots obtained from the microscopic model are shown in Figs.~\ref{fig:chaos_micro}(b) and~\ref{fig:chaos_micro}(c), respectively. 

Thus, the results presented in this section confirm the validity of the reduced mean-field Eqs.~\eqref{eq_r12v12} for describing the macroscopic dynamics of the QIF neural networks with bimodal heterogeneity. Although these equations were derived in the limit of infinite size, they predict well the averaged dynamics of a finite-size network consisting of several thousand neurons.

\section{\label{sec:Conclusions} Conclusions}

We have derived a low-dimensional system of mean field equations for a population of QIF neurons interacting via instantaneous Dirac delta pulses, when the excitability parameter has a bimodal distribution determined by a linear combination of two Lorentz functions. The reduced mean-field equations exactly describe the dynamics of the mean membrane potential and the mean spiking rate of a population in the thermodynamic limit of infinite number of neurons. The bifurcation analysis of the mean-field equations showed a rich scenario of various dynamic regimes that are not observed in a similar model with a unimodal distribution of the excitability parameter. In the latter case, the mean-field equations have only trivial fixed point attractors, and asymptotically, the neural population always approaches stationary equilibrium~\cite{Montbrio2015}. In the case of a bimodal heterogeneity, the mean field equations can have three types of attractors: a fixed point, a limit cycle, and a strange attractor. As a result, the mean membrane potential and the spiking rate of this system can asymptotically demonstrate not only stationary behavior, but also periodic and chaotic oscillations. Depending on the parameters, the system can have multiple (up to three) stable equilibrium states  characterized by different levels of the spiking rate. 
Interestingly, all oscillatory modes coexist with stable equilibrium states.

The mean-field equations are derived in the limit of an infinite network.
Nevertheless, numerical simulations of the microscopic model equations showed that they predict well the averaged dynamics of a finite-size network consisting of several thousand neurons. The advantage of the mean field equations is not only that they reduce computational costs for large-scale networks, but also allow a thorough bifurcation analysis of various dynamic regimes in the parameter space. This analysis helps to understand the mechanism of collective synchronized oscillations in the network. We have shown that two global bifurcations are responsible for the occurrence of the limit cycle oscillations: the limit point of cycle bifurcation and the saddle-node bifurcation on an invariant circle. Understanding synchronized oscillations is an important task in neuroscience. In real neural networks, synchronization can play a dual role. Under normal conditions, synchronization is responsible for cognition and learning~\cite{Singer1999,Fell2011}, while excessive synchronized oscillations are associated with malfunction in disorders such as Parkinson's disease~\cite{Hammond2007}, epilepsy~\cite{Jiruska2013,Gerster2020}, tinnitus~\cite{Tass2012tin} and others.

\section*{Acknowledgments}

This work is supported by grant No. S-MIP-21-2 of the Research Council of Lithuania. The authors are grateful to Dr. Diego Paz\'{o} for reading the manuscript and helpful comments.

\bibliographystyle{elsarticle-num}
\bibliography{references}
\end{document}